\documentclass[11pt,halfparskip]{scrartcl}

\usepackage{url}

\usepackage{mathpazo}

\usepackage{graphicx}
\usepackage[utf8]{inputenc} 
\usepackage[square]{natbib}
\usepackage{hyperref}
\usepackage{xspace}
\usepackage{booktabs}
\usepackage{fancyvrb}
\usepackage{siunitx}
\usepackage{verbatim}
\usepackage{ltablex}
\usepackage{threeparttable}
\usepackage{subcaption}

\usepackage{listings}
\usepackage{color}
\definecolor{codegreen}{rgb}{0,0.6,0}
\definecolor{codegray}{rgb}{0.5,0.5,0.5}
\definecolor{codepurple}{rgb}{0.58,0,0.82}
\definecolor{backcolour}{rgb}{0.95,0.95,0.92}
\lstdefinestyle{mystyle}{
    backgroundcolor=\color{backcolour},   
    commentstyle=\color{codegreen},
    keywordstyle=\color{magenta},
    numberstyle=\tiny\color{codegray},
    stringstyle=\color{codepurple},
    basicstyle=\scriptsize,
    breakatwhitespace=false,         
    breaklines=true,                 
    captionpos=b,                    
    keepspaces=true,                 
    numbers=left,                    
    numbersep=5pt,                  
    showspaces=false,                
    showstringspaces=false,
    showtabs=false,                  
    tabsize=2
}
\newcommand{\eg}{eg}
\begin{document}

\title{Streaming CityJSON datasets}

\author{Hugo Ledoux \and Gina Stavropoulou \and Balázs Dukai}

\date{}
\maketitle

\begin{abstract}
We introduce \emph{CityJSON Text Sequences} (CityJSONSeq in short), a format based on \emph{JSON Text Sequences} and CityJSON\@. 
CityJSONSeq was added to the CityJSON version 2.0 standard to allow us to stream very large 3D city models.
The main idea is to decompose a CityJSON dataset into its individual city objects (each building, each tree, etc.) and create several independent JSON objects of a newly defined type: \texttt{CityJSONFeature}.
We elaborate on the engineering decisions that were taken to develop CityJSONSeq, we present the open-source software we have developed to convert to and from CityJSONSeq, and we discuss different aspects of the new format, \eg\ filesize, usability, memory footprint, etc.
For several use-cases, we consider CityJSONSeq to be a better format than CityJSON because: (1) once serialised it is about 10\% more compact; (2) it takes an order of magnitude less time to process; and (3) it uses significantly less memory.
\end{abstract}



%
\section{Introduction}%
\label{sec:intro}
 
CityJSON is a JSON-based encoding for storing 3D city models that implements a subset of the CityGML data model version 3.0~\citep{OGC-CityGML3}.
As further described in \citet{19_ogdss_cityjson}, its development started in 2017 with the aim of offering an easy-to-use and web-ready alternative to the XML-encoded CityGML files~\citep{OGC-CityGML3-XML}, which in practice can be rather verbose, difficult to parse, and complex to manipulate.
The first official release of CityJSON (version 1.0) was a success: 
\begin{enumerate}
  \item its JSON-based files were on average around 7 times more compact than their CityGML-XML equivalents without loss of information. See \citet{19_ogdss_cityjson} and \url{https://www.cityjson.org/filesize/} for details, and also \citet{Praschl23} for a comparison with standard computer graphics formats; 
  \item it was adopted as an OGC Community Standard~\citep{OGC-CityJSON-v10}; 
  \item it was adopted by the Dutch government as a 3D standard to distribute nationwide 3D datasets; 
  \item several implementations and plugins have been developed, most notably FME\@.
\end{enumerate}

\begin{figure}
  \centering
  \includegraphics[width=0.6\linewidth]{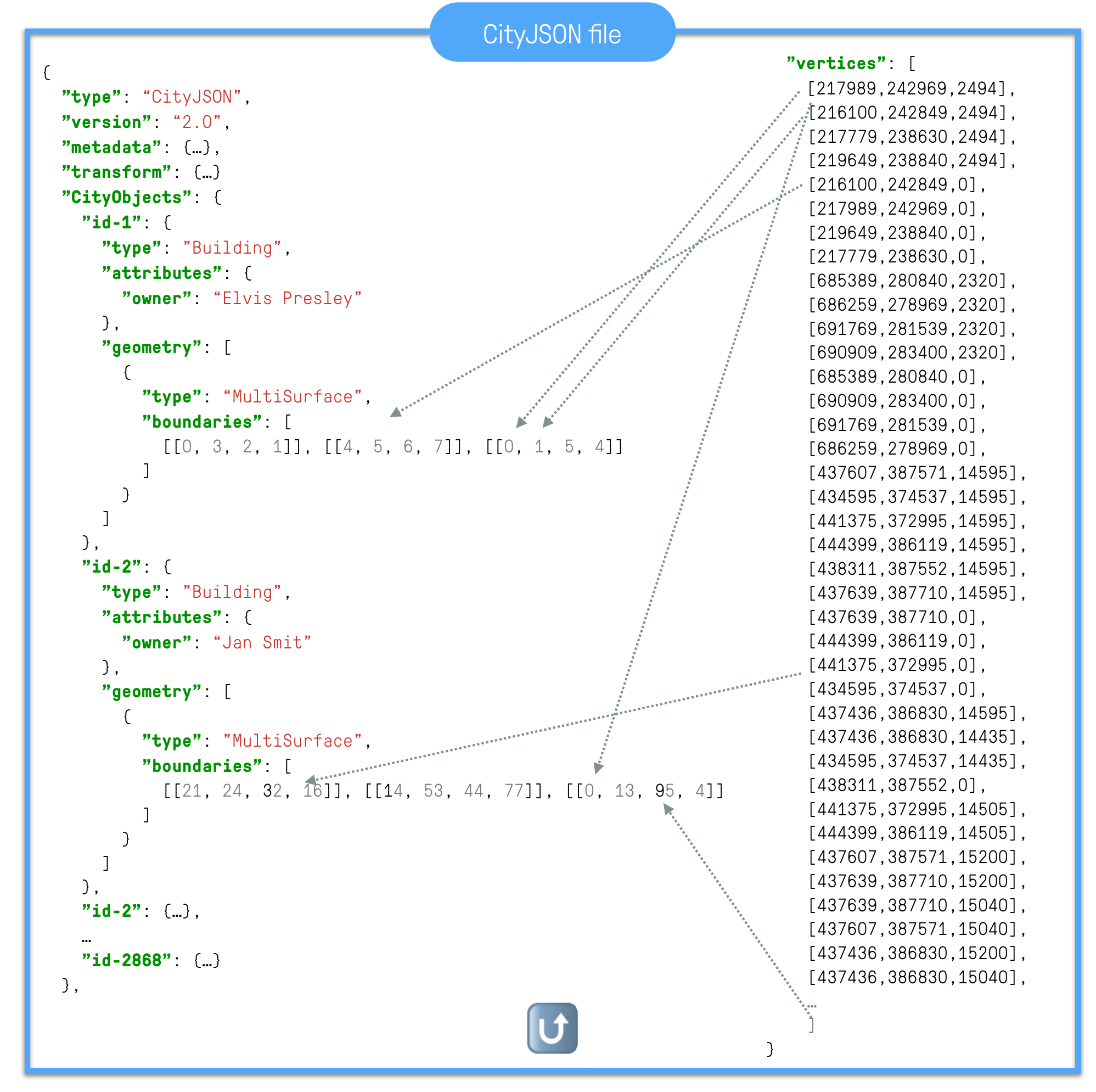}
  \caption{An example of a CityJSON file. The vertices are stored in a global list, and the position of the vertices in that list are used to represent the boundaries of the geometries (represented by the arrows, many have been left out for clarity).}%
\label{fig:cj_idea}
\end{figure}

However, the version 1.0 of CityJSON had one limitation: its structure for storing the coordinates of the geometries (see Figure~\ref{fig:cj_idea}) made the \emph{streaming} of very large datasets complex, if not impossible.
As further defined in Section~\ref{sec:streaming}, streaming refers to the possibility of downloading/transferring/processing a dataset without having to load it all in memory.
Given the increasing size of datasets of 3D cities, with CityGML-XML files often exceeding 2GB, the processing of CityJSON files has become a practical challenge.

We present in this paper \emph{CityJSON Text Sequences} (henceforth referred to as \emph{CityJSONSeq}), a format based on \emph{JSON Text Sequences}~\citep{IETF-JSONSeq} and CityJSON, and inspired by \emph{GeoJSON Text Sequences}~\citep{IETF-GeoJSONSeq}. 
CityJSONSeq was added to the CityJSON version 2.0 standard that was released in 2023 (and also standardised by the OGC, see \citet{OGC-CityJSON-v20}).
As further described in Section~\ref{sec:cityjsonseq}, the idea is to decompose a (often large) CityJSON file into its individual \emph{features} (\eg\ each \texttt{Building}, each \texttt{Bridge}, etc.) and create several JSON objects of a newly defined type \texttt{CityJSONFeature}.
These objects are either serialised into a text file, or streamed from a server to client or from one application to another.
This allows us to avoid a large list of vertices that need to be indexed, as it is the case with CityJSON files (see Section~\ref{sec:cityjson} for details).

In Section~\ref{sec:experiments}, we describe  the open-source software we have developed to convert between CityJSON and CityJSONSeq files.
We also analyse the filesizes of CityJSON and CityJSONSeq for several real-world datasets and synthetic datasets we built.
It can be observed that one advantage of CityJSONSeq, besides that files containing several thousands of features can be streamed, is that it compresses further the CityJSON files by around 12\%, sometimes more.
We discuss in Section~\ref{sec:experiments} the reasons for this interesting finding.

%
\section{Structure of a CityJSON file}%
\label{sec:cityjson}

As shown in Figure~\ref{fig:cj_idea}, a CityJSON object, which is a JSON object, represents a given geographical area, and it typically contains the following JSON properties: 
\begin{enumerate}
  \item \texttt{"type"}: it must be \texttt{"CityJSON"};
  \item \texttt{"version"}: \texttt{"2.0"} is the current version;
  \item \texttt{"metadata"}: different metadata related to the dataset can be stored. The most important is the definition of the coordinate reference system (CRS).
  \item \texttt{"transform"}: CityJSON \texttt{"vertices"} are compressed and stored as integers only. The parameters of this property allow us to convert from those integers back to real-world coordinates.
  \item \texttt{"CityObjects"}: a dictionary where the properties are the identifiers of the city objects (\emph{IDs}), which are any CityGML city object (for instance a \texttt{Building}, a \texttt{BuildingPart}, a \texttt{SolitaryVegetationObject}, etc.).
  The city objects are listed one after the other, even if some are \texttt{"children"} of others.
  As an example, for a \texttt{Building} containing 2 parts, the 3 objects will be represented at the same level and linked by their \emph{IDs}, as shown in Figure~\ref{fig:parents_children}.
  The schema is thus flat and all hierarchies have been removed.
  Each city object can have a \texttt{"parents"} and/or a \texttt{"children"} property, and this is how in the snippet the building \texttt{"id-1"} is linked to its 2 parts.
  The fact that a dictionary is used means that developers have direct access to the city objects through their IDs (and also in constant time if a hashmap is used to implement the dictionary while parsing the file).
  \item \texttt{"vertices"}: The 3D geometric primitives in CityJSON are those of the CityGML data model, which means that multi/composite solids with several parts and/or cavities are supported.
  A geometric primitive does not list all the coordinates of its vertices, instead the coordinates of the vertices are stored in a separate array (the \texttt{"vertices"} property of the CityJSON object), and the geometric primitives refer to the position of a vertex in that array.
  This indexing mechanism has been successfully used for many years by the computer graphics community in formats as \emph{Wavefront OBJ}\footnote{\url{https://en.wikipedia.org/wiki/Wavefront_.obj_file}}.
  There are several advantages to this approach.
  First, the files can be compressed: 3D vertices are often shared by several surfaces, and repeating them can be costly, especially if they are very precise (sub-millimetre precision is often used).
  Second, this approach increases the topological relationships that are explicitly stored in the file, and several operations (\eg\ determining building adjacency) can be sped up and made more robust.
  Third, it is very easy to convert all coordinates to a representation listing; the inverse is not true. 
  However, this list of vertices is the reason why the streaming of geometries is problematic, since in practice it can contain several millions vertices.
  To be able to reconstruct a single Building, all the \texttt{"vertices"} need to loaded in memory, which can mean waiting for millions of unused vertices to be deserialised.
  \item \texttt{"appearances"}: Both textures and materials for surfaces are supported.  
  The material of a surface is represented with the X3D specifications\footnote{\url{https://en.wikipedia.org/wiki/X3D}}, and for the textures the COLLADA specifications\footnote{\url{https://www.khronos.org/collada/}} are reused.
  \item \texttt{"geometry-templates"}: 
  Geometry templates are geometries defined once and reused by applying a translation, a rotation, and/or a scaling.
  They are mostly used for city objects like trees, bus stops, and lamp posts. 
\end{enumerate}

\begin{figure}
  \centering
\begin{lstlisting}
  "CityObjects": {
    "id-1": {
      "type": "Building",
      "attributes": {...},
      "children": ["id-2", "id-3"],
      "geometry": [{...}]
    },
    "id-2": {
      "type": "BuildingPart",
      "parents": ["id-1"],
      "geometry": [{...}]
      ...
    },
    "id-3": {
      "type": "BuildingPart",
      "parents": ["id-1"],
      "geometry": [{...}]
      ...
    }
    ...
    "id-77": {}
  }
\end{lstlisting}
  \caption{CityJSON mechanism to flatten out the schema: the city objects are stored in a flat list, and they are linked together with the properties \texttt{"parents"} and \texttt{"children"}.}%
\label{fig:parents_children}
\end{figure}

%
\section{Streaming (3D) datasets}%
\label{sec:streaming}
\begin{figure*}[th]
  \centering
  \includegraphics[width=0.95\linewidth]{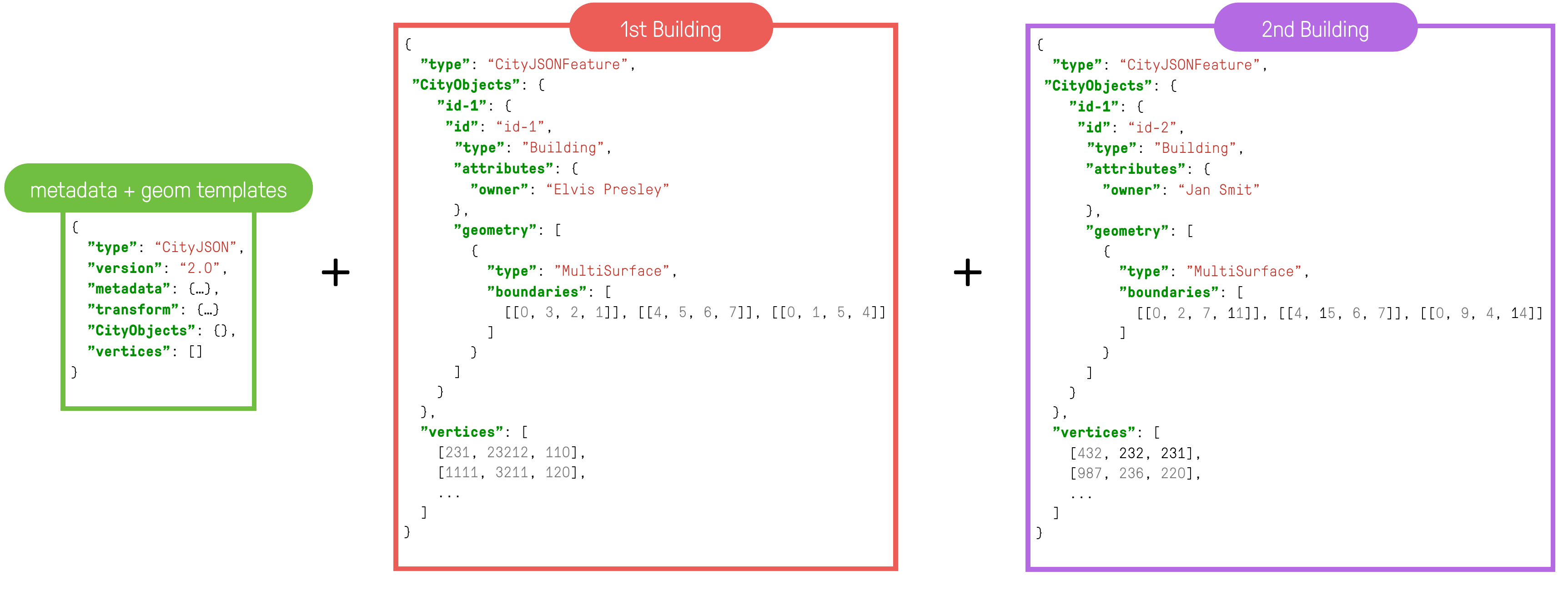}
  \caption{The CityJSONSeq of a CityJSON dataset with two buildings contains three JSON objects: one for the metadata, plus one for each building.}%
\label{fig:cjseq_idea}
\end{figure*}
In the context of geo-information, a \emph{stream} is a sequence of data that is available over a period of time, and ``can be thought of as items on a conveyor belt being processed one at a time rather than in large batches''\footnote{From \url{https://en.wikipedia.org/wiki/Stream_(computing)}}.

For GML-based formats (which are feature-centric), modifying a file for the purpose of streaming is usually a simple task that involves sending the features in the dataset one-by-one.
However, notice that this is only true for GML files that follow the Simple Feature paradigm~\citep{OGC-SF}.
For more complex data models like CityGML, where geometry templates and \emph{XLinks} are used, streaming often requires a large amount of pre-processing.
\emph{XLinks} are links between elements in a file (similar to \emph{pointers} in a computer program), a concrete example in a CityGML file is a cube that lists the geometries of 5 of its surfaces, but its 6th surface is simply a link to another surface somewhere else in the file (which belongs to another building for instance).
If the linked surface has not be seen  in the stream yet, the receiver cannot process the cube and needs to wait for that specific 6th surface to appear (\eg\ to calculate its volume).
It is not always possible to (re-)order a CityGML file so that all the references can be resolved without having to store extra information until it appears in the stream.
However, it is always possible to resolve the \emph{XLinks} before streaming a dataset (that is, in our example, copy the geometry of the 6th surface to the cube), but this means that the filesize will increase, and that converting back to the original file will not be possible.

For the \emph{GeoJSON} format~\citep{IETF-GeoJSON}, which follows the Simple Feature paradigm, creating a stream is trivial since each of the features in the dataset becomes one JSON object serialised to one line~\citep{IETF-GeoJSONSeq}.
There are no links possible between features, each JSON object is independent.

For formats that use a global indexing of vertices, such as CityJSON and most formats used for storing meshes in computer graphics (\eg\ OBJ and STL), the reorganisation of the elements in a file is more complex but nonetheless possible.
\citet{Isenburg03} describe algorithms and tools that interleave the vertices and faces in a file (instead of having one large list of vertices at the end) and add simple tags to inform that specific vertices are not used anymore in the stream (and thus can be freed from memory).
This allows us, in theory, to process/edit/manipulate infinitely large meshes, since they never have to be completely loaded in memory.
The idea is exemplified by the construction of gridded terrains that are gigabytes in size~\citep{Isenburg06-1}.
However, this cannot be implemented in CityJSON directly because all the vertices need to be listed in the JSON property \texttt{"vertices"}.

%
\section{CityJSON Text Sequences}%
\label{sec:cityjsonseq}

As shown in Figure~\ref{fig:cjseq_idea}, a CityJSONSeq decomposes a CityJSON object into its features to create a sequence of several JSON objects.
Those JSON objects are of type \texttt{CityJSONFeature}, which allows the storage of a single feature, for instance a \texttt{Building}, together with its ``children'' objects (\eg\ a \texttt{BuildingPart} and/or a \texttt{BuildingInstallation}). 
Each feature is independent, it has its own list of vertices (which is thus \emph{local} to the JSON object, and is usually rather small, see next section for details) and its own textures and materials (if any).
The allowed properties are shown in Figure~\ref{fig:cjfeature}; notice that the \texttt{"id"} property is used to clearly identity the ``parent'' of the feature, in case there are children.
\begin{figure}
\begin{lstlisting}
{
  "type": "CityJSONFeature",
  "id": "id-1", 
  "CityObjects": {
    "id-1": {
      "type": "Building", 
      "attributes": { 
        "roofType": "gabled roof"
      },
      "children": ["mybalcony"],
      "geometry": [...]
    },
    "mybalcony": {
      "type": "BuildingInstallation", 
      "parents": ["id-1"],
      "geometry": [...]
    }
  },
  "appearance": {...}
  "vertices": [...]
}
\end{lstlisting}
\caption{An example of a \texttt{CityJSONFeature} for a Building with a balcony referenced in its \texttt{"children"} property.}%
\label{fig:cjfeature}
\end{figure}

CityJSONSeq follows the specifications of \texttt{ndjson} (newline delimited JSON)\footnote{\url{https://github.com/ndjson/ndjson-spec/}} and two constraints are added for handling CityJSON\@:
\begin{enumerate}
  \item each JSON Object must conform to the \emph{JSON Data Interchange Format specifications}~\citep{IETF-JSON} and be written as a UTF-8 string;
  \item each JSON Object must be followed by a new-line (LF: \texttt{"\textbackslash n"}) character, and it may be preceded by a carriage-return (CR: \texttt{"\textbackslash r"});
  \item a JSON Object must not contain the new-line or carriage-return characters;
  \item the first JSON Object must be of type \texttt{CityJSON};
  \item the following JSON Objects must be of type \texttt{CityJSONFeature}.
\end{enumerate}

Note that a \texttt{CityJSONFeature} object does not contain all the information that is required for reconstructing the feature. 
Most commonly, the \texttt{"transform"} property, the CRS, and the \texttt{"geometry-templates"} must be known in order to correctly reconstruct and georeference the city objects. 
The rule \#4 ensures that those are available. 
The CityJSON object must contain a \texttt{"transform"} property and eventually the other properties if needed; \texttt{"CityObjects"} and \texttt{"vertices"} must be present but they must be empty (to ensure that the JSON object is valid).

The CityJSONSeq for Figure~\ref{fig:cjseq_idea} is shown in Figure~\ref{fig:stream}.
\begin{figure*}
  \begin{lstlisting}
  {"type":"CityJSON","version":"2.0","transform": {"scale":[1.0,1.0,1.0],"translate":[0.0, 0.0, 0.0]},"metadata":{"referenceSystem":"https://www.opengis.net/def/crs/EPSG/0/7415"},"CityObjects":{},"vertices":[]}
  {"type":"CityJSONFeature","id":"id-1","CityObjects":{...},"vertices":[...]} 
  {"type":"CityJSONFeature","id":"id-2","CityObjects":{...},"vertices":[...]} 
  \end{lstlisting}
\caption{An example of a CityJSONSeq stream containing 3 features.}%
\label{fig:stream}
\end{figure*}

%
\section{Experiments with real-world datasets}%
\label{sec:experiments}

To convert between CityJSON and CityJSONSeq files (and vice-versa), we have developed the open-source software \emph{cjseq}, which is available at \url{https://github.com/cityjson/cjseq/} under a permissive open-source license.
The command-line program handles the conversion not only of the geometries, but also of the materials, the textures, and the geometry templates that the dataset could contain.
It includes three sub-commands: 
\begin{enumerate}
  \item \texttt{cat}: CityJSON $\rightarrow$ CityJSONSeq;
  \item \texttt{collect}: CityJSONSeq $\rightarrow$ CityJSON;
  \item \texttt{filter}: to filter city objects in a CityJSONSeq, randomly or based on a bounding box.
\end{enumerate}

It should be observed that the conversion is an efficient process: the rather large dataset \emph{Helskinki} from Table~\ref{tab:datasets}, which contains more than \qty{77000} buildings and whose CityJSON file is \qty{572}{\mega\byte}, takes only \qty{4.7}{\sec} to be converted to a CityJSONSeq file, and the reverse operation takes \qty{5.7}{\sec} (on a standard laptop).

\subsection{Filesize comparison}

\begin{table}
  \centering
  \begin{threeparttable}
  \caption{The datasets used for the benchmark. }%
  \label{tab:datasets}
  \footnotesize
  \begin{tabular}
    {@{}lcccccrrrcrrr@{}}\toprule
    &&  \multicolumn{2}{c}{\textbf{dataset}} && \multicolumn{3}{c}{\textbf{size of file}} && \multicolumn{3}{c}{\textbf{vertices}}   \\ 
    \cmidrule{3-4} \cmidrule{6-8} \cmidrule{10-12} 
     && CityObjects &  app.\footnotesize ${}^{\text{(a)}}$ && CityJSON & CityJSONSeq & compr.\footnotesize ${}^{\text{(b)}}$ && total & largest\footnotesize ${}^{\text{(c)}}$ & shared\footnotesize ${}^{\text{(d)}}$ \\
    \midrule
    \textbf{3DBAG}          && \qty{1110} bldgs    &         && \qty{6.7}{\mega\byte} & \qty{5.9}{\mega\byte} & 12\%  &&     \num{82509} &    \num{4112} &  0.1\% \\
    \textbf{3DBV}           && \qty{71634} misc   &         && \qty{378}{\mega\byte} & \qty{317}{\mega\byte} & 16\%  &&   \num{4110319} &  \num{116670} & 21.0\% \\
    \textbf{Helsinki}       && \qty{77231} bldgs   &         && \qty{572}{\mega\byte} & \qty{412}{\mega\byte} & 28\%  &&   \num{3038576} &    \num{2202} &  0.0\% \\
    \textbf{Helsinki\_tex}  && \qty{77231} bldgs   & tex     && \qty{713}{\mega\byte} & \qty{644}{\mega\byte} & 10\%  &&   \num{3038576} &    \num{2202} &  0.0\% \\
    \textbf{Ingolstadt}     && \qty{55} bldgs      &         && \qty{4.8}{\mega\byte} & \qty{3.8}{\mega\byte} & 25\%  &&     \num{87972} &   \num{12800} &  0.0\% \\
    \textbf{Montréal}       && \qty{294} bldgs     & tex     && \qty{5.4}{\mega\byte} & \qty{4.6}{\mega\byte} & 15\%  &&     \num{31585} &    \num{3393} &  2.0\% \\
    \textbf{NYC}            && \qty{23777} bldgs   &         && \qty{105}{\mega\byte} &  \qty{95}{\mega\byte} & 10\%  &&   \num{1035804} &    \num{2608} &  0.8\% \\
    \textbf{Railway}        && \qty{50} misc      & tex+mat && \qty{4.3}{\mega\byte} & \qty{4.0}{\mega\byte} &  8\%  &&     \num{73554} &   \num{14966} &  0.4\% \\
    \textbf{Rotterdam}      && \qty{853} bldgs     & tex     && \qty{2.6}{\mega\byte} & \qty{2.7}{\mega\byte} & -4\%  &&     \num{22246} &     \num{631} & 20.0\% \\
    \textbf{Vienna}         && \qty{307} bldgs     &         && \qty{5.4}{\mega\byte} & \qty{4.8}{\mega\byte} & 11\%  &&     \num{47220} &    \num{2025} &  0.0\% \\
    \textbf{Zürich}         && \qty{52834} bldgs   &         && \qty{279}{\mega\byte} & \qty{247}{\mega\byte} & 11\%  &&   \num{3472989} &    \num{4069} &  2.6\% \\
    \bottomrule
  \end{tabular}
    \begin{tablenotes}[flushleft]
      \footnotesize
      \item ${}^{\text{(a)}}$ appearance: `tex' is textures stored; `mat' is material stored
      \item ${}^{\text{(b)}}$ compressi{}on factor is $\frac{size(CityJSON) - size(CityJSONSeq)}{size(CityJSON)}$
      \item ${}^{\text{(c)}}$ number of vertices in the largest feature of the stream
      \item ${}^{\text{(d)}}$ percentage of vertices that are used to represent different city objects
    \end{tablenotes}
  \end{threeparttable}
\end{table}

We have converted with \emph{cjseq} several publicly available files, and Table~\ref{tab:datasets} shows an overview of the files stored both in CityJSON and CityJSONSeq.
The files are available in the reproducibility repository of the paper\footnote{\url{https://github.com/cityjson/paper_cjseq}}.

First observe that---contrary to intuition---the filesize of a dataset serialised as a CityJSONSeq file is around 12\% compacter than serialised as a CityJSON file, and in the case of \emph{Helsinki} it is 28\%.
An even larger compression factor is noted in most datasets whose texture, materials, semantics and attributes have been removed. 
The main reason for this is that the indices of the vertices are low integers for each feature (because the lowest index in each feature is always ``0'' and is incremented by 1 until the total number of vertices), and they do not increase to very large integers in contrast to the vertices in CityJSON\@.
For instance, the dataset \emph{Helsinki} contains a total of more than 3 millions vertices, but its largest feature contains only but 2202 vertices.
The fact that many indices are used for representing the geometries (and the textures) means that if several large numbers are used then the filesize will grow; if the maximum vertex index is around 2000 for each feature then the filesize will be reduced.

Only one dataset sees its filesize slightly increase, by 4\%, when serialised to a CityJSONSeq file: \emph{Rotterdam}.
The reasons for the increase (or decrease) are many, and we discuss in the following the 3 most relevant: (1) the total number of vertices; (2) the number of shared vertices; (3) the presence of textures.

\begin{figure}
  \centering
  \begin{subfigure}[b]{0.5\linewidth}
    \centering
    \includegraphics[width=\textwidth]{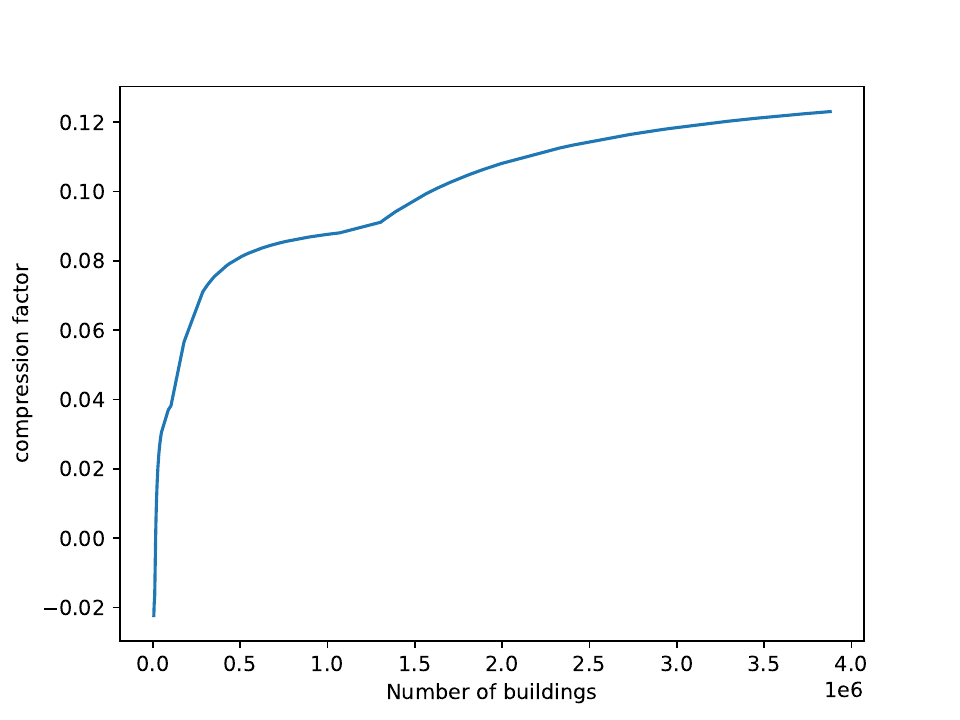}
    \caption{}%
  \label{fig:compression_random}
  \end{subfigure}%
  \begin{subfigure}[b]{0.5\linewidth}
    \centering
    \includegraphics[width=\textwidth]{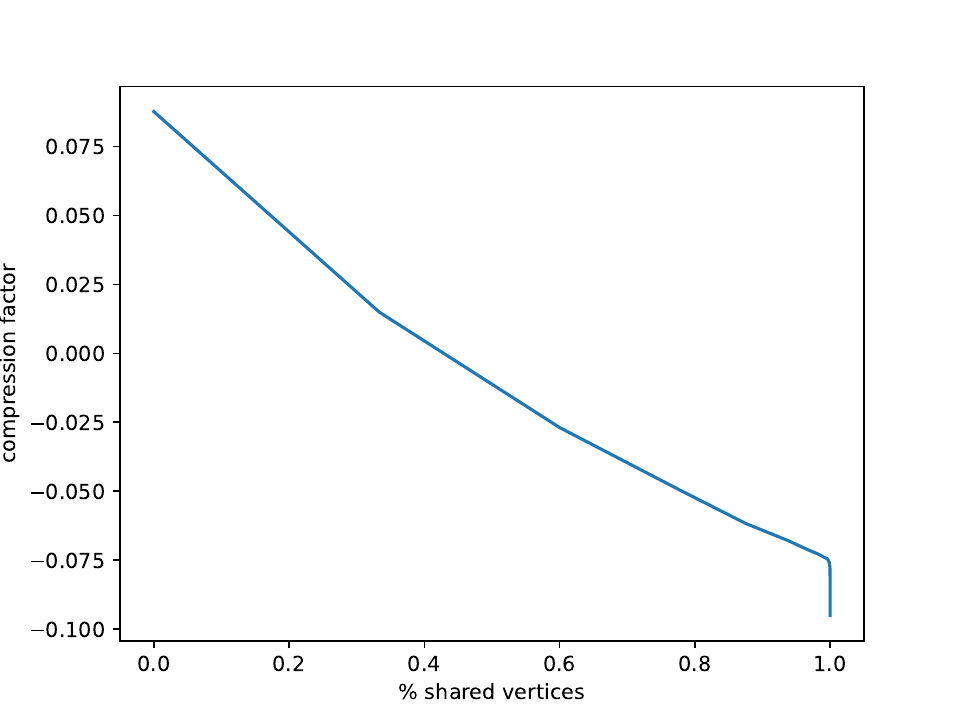}
    \caption{}%
  \label{fig:compression_adjacent}
  \end{subfigure}
  \caption{Compression factor of CityJSONSeq files for different synthetic datasets. \textbf{(a)} Based on the number of buildings; the buildings are stored as simple cuboids that are randomly generated. \textbf{(b)} Based on the percentage of shared vertices; the same cuboid buildings are used, but we position them adjacent to each others to create shared vertices.}%
\label{fig:compression}
\end{figure}

\paragraph{Number of vertices.} 
If a dataset has few vertices, as it is the case with \emph{Rotterdam}, then the indices will not be large integers and this might not be favourable for the compression.
As an experiment, we have created around 100 synthetic CityJSON datasets containing buildings, and each building is represented as a simple cube, which is randomly generated.
There are no attributes, no semantics, and no textures/materials. 
Figure~\ref{fig:compression_random} shows that, as the CityJSON filesize increases, the compression factor increases.
The smallest file contains only 526 buildings and its compression factor is -2\% (thus CityJSONSeq has a larger filesize than that of CityJSON), while the largest file has \num{3960105} buildings, and a compression factor of more than 12\%.

\paragraph{Shared vertices.} 
The number of shared vertices between different city objects also influences the compression factor. 
Shared vertices are those used to represent walls incident to two adjacent buildings.
In CityJSON they are conceptually the same vertices and each of the surfaces refer to them, but in CityJSONSeq they have to be listed separately in each of the buildings.
It should be said that most of the datasets have very few vertices that are shared (most have less than 2\%, except 2 datasets have around 20\%, \emph{Rotterdam} being one of them).
To understand the correlation between the compression factor and the percentage of shared vertices in a datasets, we have modified the script to generate random cuboid buildings: the distribution of the buildings is not random, we have enforced that several buildings are adjacent to others (so that they share vertices with other buildings).
The relationship between the compression and the percentage of shared vertices can be seen in Figure~\ref{fig:compression_adjacent} for around 100 datasets containing exactly \num{1000000} buildings.
If the number of shared vertices is 0\% this means that we have \num{1000000} buildings that are disconnected; in this case we obtain a compression factor of around 8\% (as was the case in Figure~\ref{fig:compression_random}).
If all the buildings are adjacent to another one (thus nearly 100\% of the vertices are shared), then we can see that the compression factor is about -10\% (which means that the size of the CityJSONSeq file is larger than that of the CityJSON file).

\paragraph{Textures.} 
It should also be noticed that the attributes attached to city objects, as well as the semantics attached to surfaces, have no influence on the compression factor since they are local to each city object.
However, we can state that textures have an influence on the compression factor.
See for instance the dataset \emph{Helsinki} and its counterpart \emph{Helsinki\_tex} (which is the same the same geometries and attributes, only the textures were removed).
The dataset with textures has a compression of 10\% while the one without 28\%.
This is explained by the fact that the \texttt{"textures"} property must be used for each feature, while in a CityJSON object they are all stored at only one location.
Since textures can be used by several features (all the bricks of a building could use the same one), this means that often the same properties for textures are copied to several features.

\subsection{Processing speed comparison}

\begin{table}
  \centering
  \caption{Comparison of the processing time and maximum RAM usage for processing CityJSON and CityJSONSeq files. The resident set size (RSS) is used, which is the portion of main memory occupied by the Python script.}
  \small
  \begin{tabular}
    {@{}lcrrcrrr@{}}\toprule
    &&  \multicolumn{2}{c}{\textbf{RAM used (MB)}} && \multicolumn{3}{c}{\textbf{time (s)}} \\ 
    \cmidrule{3-4} \cmidrule{6-8} 
     && CityJSON & CityJSONSeq && CityJSON & CityJSONSeq & diff \\
    \midrule
     \textbf{3DBAG}         &&   76.9 &  16.1  &&   0.10 & 0.07 & 1.4X \\
     \textbf{3DBV}          && 4101.8 & 123.8  &&  10.95 & 3.59 & 3.1X \\
     \textbf{Helsinki}      && 3743.1 &  15.0  &&  13.39 & 2.74 & 4.9X \\
     \textbf{Helsinki\_tex} && 5004.8 &  19.1  &&  29.60 & 4.72 & 6.3X \\
     \textbf{Ingolstadt}    &&   65.5 &  21.3  &&   0.08 & 0.06 & 1.3X \\
     \textbf{Montréal}      &&   79.3 &  20.8  &&   0.11 & 0.07 & 1.6X \\
     \textbf{NYC}           &&  949.5 &  16.0  &&   1.78 & 0.70 & 2.5X \\
     \textbf{Railway}       &&   69.6 &  29.6  &&   0.09 & 0.07 & 1.3X \\
     \textbf{Rotterdam}     &&   42.4 &  14.6  &&   0.04 & 0.04 & 1.0X \\
     \textbf{Vienna}        &&   60.1 &  15.7  &&   0.06 & 0.05 & 1.2X \\
     \textbf{Zurich}        && 2793.1 &  16.3  &&   6.05 & 2.00 & 3.0X \\
    \bottomrule
  \end{tabular}%
  \label{tab:ramtime}
\end{table}

We compare the speed and memory footprint of accessing each city object in a dataset.
This operation is common in applications that manipulate 3D city models.

When a city model is stored in its entirety in one CityJSON object, we need to deserialise the whole CityJSON object into memory in order to access the \texttt{"transform"} and \texttt{"vertices"} properties for instance.

With a CityJSONSeq file, we can read the file line by line, processing and discarding the city objects one by one (and thus never have in memory more than the city object itself and the first JSON object in the stream).
As shown in the experiments below, this allows for very efficient operations in terms of both CPU and memory usage.

%

We have processed all the datasets from Table~\ref{tab:datasets} with two simple Python scripts that iterate through the city objects and their geometries, and increment a global counter for the geometry type (\texttt{Solid} or \texttt{MultiSurface}) and report it at the end.
This is just an example of a simple local operation, any other operation such as calculating the area of the façades or counting the number of windows could have been performed.
The results for both the maximum memory footprint and the time used are shown in Table~\ref{tab:ramtime}.
The scripts are available in the reproducibility repository of the paper\footnote{\url{https://github.com/cityjson/paper_cjseq}}.

Notice that the dataset \emph{3DBV} contains not only buildings but also the terrain, and a few large areas are stored as a triangulation containing a very large amount of vertices; this is the reason why the maximum memory use is larger than for other datasets.

The results in Table~\ref{tab:ramtime} indicate that there is a significant benefit to using CityJSONSeq over CityJSON, at least for operations that do not require analysing or processing city objects that are close to each other.
Operations like calculating volumes, merging and subsetting files, finding city objects with specific attributes, etc.\ will all use significantly less memory and will be significantly faster.
Operations like modifying the CRS or updating the metadata would not even require to loop through the features, just to alter the first object in the file.
Operations like calculating the surface of shared walls~\citep{Agugiaro22} are however not suitable for streams.

%
\section{Discussion and future work}%
\label{sec:discussion}

While CityJSONSeq was developed mostly for streaming large 3D city datasets, the fact that it has a much lower memory footprint and that it takes an order of magnitude less time to process (for some local operations) makes it an attractive alternative to CityJSON for several use-cases.

It should be noticed that the CityJSON specification does not prescribe the storage of CityJSONSeq, only the structure of a CityJSONSeq stream.
In practice, CityJSONSeq can be stored in a variety of ways, for instance in a single file, each feature in a separate file, in a database, etc.
The optimal storage solution depends on the implementing application.
As a concrete example, CityJSONSeq is used by \emph{cjdb}~\citep{Powalka23}, an importer/exporter tool that stores one feature per row in PostgresSQL.
Additionally, in order to facilitate pagination, CityJSONSeq is the return format of the 3DBAG API\footnote{\url{https://api.3dbag.nl/}}, which contains all 10 million buildings in the Netherlands with detailed roofs~\citep{Peters22}.

From the point-of-view of practitioners, we should stress that CityJSONSeq can easily be processed with \emph{Unix pipes} (also called \emph{pipelines}).
Pipelines allow us to chain several processes together, the output of a process becomes the input of the next one, and so on. 
Given 2 processes, the 2nd one can usually start before the 1st one has finished processing all the data.
This means that, in practice, the processing of a dataset can be performed by writing several programs that perform a few small tasks; this makes the development and maintenance of code simpler.
Moreover, it allows us to write the code in different languages. 
The pipelines used for preparing the datasets used in this paper were actually a mix of Python (the program \emph{cjio} and \emph{cjdb}) and Rust (\emph{cjseq} and \emph{cjval}), and another program written in C++ could easy be added.

Finally, because the structure of CityJSONSeq is nearly the same as that of CityJSON, in practice adding support for CityJSONSeq requires a minimum amount of effort since most of the code to parse and/or generate CityJSON objects can be reused. 
We have already added support for CityJSONSeq in a few of the open-source tools\footnote{\url{https://www.cityjson.org/software/}} and we will continue in the future.

\bibliographystyle{abbrvnat}
\bibliography{refs}

\end{document}